%% file: data_testing.tex
\documentclass[sigconf]{acmart}

\usepackage{framed}
\usepackage{balance}
\balance
\usepackage{enumitem}
\usepackage[most]{tcolorbox}
\usepackage{placeins}
\usepackage{color}
\usepackage{ifthen}
\newboolean{showcomments}
\setboolean{showcomments}{true} 
\ifthenelse{\boolean{showcomments}}
    {
        \newcommand{\emelie}[1]{\textcolor{red}{{\it [Emelie says: #1]}}}
        \newcommand{\markus}[1]{\textcolor{blue}{{\it [Markus says: #1]}}}
        \newcommand{\song}[1]{\textcolor{olive}{{\it [Song says: #1]}}}
        \newcommand{\story}[1]{\textcolor{orange}{{\it [Story: #1]}}}
        \newcommand{\sergio}[1]{\textcolor{brown}{{\it [Sergio: #1]}}}
    }
    {
        \newcommand{\emelie}[1]{}
        \newcommand{\markus}[1]{}
        \newcommand{\song}[1]{}
        \newcommand{\story}[1]{}
        \newcommand{\sergio}[1]{}
    }
    
\begin{document}
\title{Exploring ML testing in practice -- Lessons learned from an interactive rapid review with Axis Communications}

\author{Qunying Song}
\orcid{XXX}
\affiliation{%
  \institution{Lund University}
  \city{Lund}
  \country{Sweden}
}
\email{qunying.song@cs.lth.se}

\author{Markus Borg}
\orcid{XXX}
\affiliation{%
  \institution{RISE Research Institutes of Sweden and Lund University}
  \city{Lund}
  \country{Sweden}
}
\email{markus.borg@ri.se}

\author{Emelie Engström}
\orcid{XXX}
\affiliation{%
  \institution{Lund University}
  \city{Lund}
  \country{Sweden}
}
\email{emelie.engstrom@cs.lth.se}

\author{Håkan Ardö}
\orcid{XXX}
\affiliation{%
  \institution{Axis Communications}
  \city{Lund}
  \country{Sweden}
}
\email{hakan.x.ardo@axis.com}

\author{Sergio Rico}
\orcid{XXX}
\affiliation{%
  \institution{Lund University}
  \city{Lund}
  \country{Sweden}
}
\email{sergio.rico@cs.lth.se}

\renewcommand{\shortauthors}{Song et al.}

\begin{abstract}
There is a growing interest in industry and academia in machine learning (ML) testing. We believe that industry and academia need to learn together to produce rigorous and relevant knowledge. In this study, we initiate a collaboration between stakeholders from one case company, one research institute, and one university. To establish a common view of the problem domain, we applied an interactive rapid review of the state of the art. Four researchers from Lund University and RISE Research Institutes and four practitioners from Axis Communications reviewed a set of 180 primary studies on ML testing. We developed a taxonomy for the communication around ML testing challenges and results and identified a list of 12 review questions relevant for Axis Communications. The three most important questions (data testing, metrics for assessment, and test generation) were mapped to the literature, and an in-depth analysis of the 35 primary studies matching the most important question (data testing) was made. A final set of the five best matches were analysed and we reflect on the criteria for applicability and relevance for the industry. The taxonomies are helpful for communication but not final. Furthermore, there was no perfect match to the case company's investigated review question (data testing). However, we extracted relevant approaches from the five studies on a conceptual level to support later context-specific improvements. We found the interactive rapid review approach useful for triggering and aligning communication between the different stakeholders. 
\end{abstract}

\copyrightyear{2022} 
\acmYear{2022} 
\acmConference[CAIN '22]{International Conference on AI Engineering}{May 22--23, 2022}{Pittsburgh, USA}

%
%



\maketitle

\input{body.tex}

\balance
\bibliographystyle{ACM-Reference-Format}
\bibliography{sim_refs}

\end{document}

%% file: body.tex
\section{Introduction} \label{sec:intro}
Artificial intelligence (AI) applications have grown in popularity and pervasiveness.
Among the AI applications currently in use, machine learning (ML) is the dominant technique with active communities in academia and industry~\cite{LWAKATARE2020106368}.
Enterprises across diverse industry domains want to harness the new possibilities promoted by ML. 
However, due to their impact and increasing use in safety-critical domains, we need to develop ways to build trust in these applications.
Bosch \textit{et al.} calls for increased research on \textit{AI engineering}~\cite{bosch2021engineering}, i.e., an evolution of software engineering practices and processes to meet the needs of systems development that incorporate trained ML models.
These systems, in contrast to most traditional systems, have a probabilistic behavior~\cite{8718214}. Therefore, we need new approaches and solutions or adapt the existing solutions to new challenges~\cite{borg2019safely,8804457}.

In this paper, we focus on ML testing, emphasizing applications of ML-based computer vision.
ML testing has been a popular research topic in the last few years. Secondary studies show a rapidly increasing publication trend~\cite{zhang2020machine,riccio2020testing,sherin2019systematic} and dedicated academic workshops and conferences have been established.
As novel ML testing results are constantly published, both researchers and practitioners need ways to organize the information and sift through the massive academic output. Furthermore, there is a need for effective ways to match research proposals with application-specific industry needs~\cite{carleton2020expert}.


An approach helpful in the inception of a collaborative project is interactive rapid reviews (IRRs)~\cite{Rico2020GuidelinesFC}. 
An IRR is a collaborative effort between researchers and practitioners that aims to identify and synthesize relevant research outcomes for the practitioners in their context.
Apparently, an IRR could be a beneficial tool for new collaborative projects to explore interests, facilitate the exchange of ideas, and promote mutual understanding.


We conducted an IRR on ML testing with Axis Communications (hereafter Axis). The long-term goal of the IRR was to initiate a collaboration on ML testing between researchers from Lund University, RISE Research Institutes of Sweden, and practitioners at Axis. As a means to that end, and a short-term goal, the IRR should identify the solution proposals from the academic community that are the most likely to provide value for Axis. 


The contributions of this paper are the following. First, we developed a taxonomy about practical challenges and available research results on ML testing that helped us learn about the domain and navigate the research results (Sec. 4.1). Second, we compiled a list of twelve practical challenges, identified during the IRR at Axis, related to ML testing (Sec. 4.2). Third, we proposed a preliminary mapping, i.e, a potential connection between research results and practical challenges for three prioritized challenges at Axis (Sec4.3). Finally, we conducted an in-depth review of the 35 primary studies mapped to the highest priority topic, i.e., ``How to test the dataset?'' We extracted nine technological rules and identified context factors impacting the application of ML testing solutions found in the academic sources (Sec. 4.5).  


\section{Background and related work} \label{sec:bg}
This section presents our position on AI quality and its connection to ML testing. Moreover, we introduce IRRs and the industrial case context.  

\subsection{AI quality and ML Testing}
Quality is a multi-faceted concept that is notoriously difficult to nail down. Adding AI on top of this further exacerbates the challenge. Still, we posit that AI quality is going to be an increasingly important concept to ensure the trustworthiness that future AI systems must provide. AIQ is a regional effort to gather interested parties on AI quality, with a particular focus on the subset of AI that realizes functionality through supervised or unsupervised machine learning, i.e., MLware. 

We adhere to the definition of AI quality as ``the capability of MLware to satisfy stated and implied needs under specified conditions while the underlying data satisfy the requirements specific to the application and its context’’~\cite{borg_AIQ_2021}. The definition stresses that MLware combines data and conventional source code; thus, its quality is defined as an amalgamation of corresponding quality definitions from the IEC/ISO~25000 series~\cite{iso25010,iso25012}. The definition is in line with discussions by Felderer \textit{et al.} in the context of testing data-intensive systems~\cite{felderer2019testing}. Moreover, the emphasis on data quality assurance is central in this paper.

Inspired by Bjarnason \textit{et al.}'s work on requirements engineering (RE) and software testing~\cite{bjarnason2014challenges}, our position is that AI quality assurance must be tackled from two directions. RE and testing must support MLware development projects as two bookends. As MLware is sensitive to changes, as Sculley \emph{et al.} put it ``changing anything changes everything’’~\cite{sculley_hidden_2015}, aligning RE and testing is perhaps even more important than for conventional software engineering. Within AIQ, we have addressed RE for ML~\cite{vogelsang_requirements_2019,borg2021exploring}, ML testing~\cite{borg2021test,moghadam2021deeper,ebadi2021efficient}, and MLOps from the perspective of alignment~\cite{borg2022agility}. In this paper, we again focus on ML testing.

ML testing is a rapidly growing research area that evolves software testing to meet the novel characteristics of ML-based systems.  
We used three secondary studies of ML testing~\cite{zhang2020machine, riccio2020testing, sherin2019systematic} as the basis for the current work. The three secondary studies were considered the latest in the field when we initiated the current study. Given that the area ML testing is fairly new, we did not expect very old publications. However, we did not explicitly exclude them either. The endpoint in the range was when we started the work, and we included all secondary studies we were aware of. Among the secondary studies we selected, two were published in 2020~\cite{zhang2020machine, riccio2020testing} and the other one in 2019~\cite{sherin2019systematic}. They used a systematic approach for searching, extracting, and synthesizing relevant studies on the topic of ML testing. It is also worth noting that Zhang \emph{et al.}~\cite{zhang2020machine} and Ricco \emph{et al.}~\cite{riccio2020testing} have also included arXiv pre-prints to be more extensive, and have identified 138 and 70 primary studies, respectively. In contrast, Sherin \emph{et al.} restricted the literature search to peer-reviewed publications only and have identified 37 papers in their study~\cite{sherin2019systematic}. In total, we have collected 180 unique primary studies based on the three secondary studies.

The three secondary studies report an increasing number of ML testing papers in recent years. The trend is the increased use of ML in various application domains and the importance of techniques for testing such applications. This is particularly evident as ML-based applications are deployed in both safety-critical and mission-critical contexts. Specifically, the majority of the studies that have been surveyed in Zhang \emph{et al.}~\cite{zhang2020machine} are focusing on testing the correctness and robustness of supervised machine learning systems, while other type of learning such as unsupervised learning and reinforcement learning, and testing perspectives such as interpretability, efficiency, or privacy are much less studied. The analysis is consistent with the observations from Sherin \emph{et al.}~\cite{sherin2019systematic} that further attention is required to test the non-functional perspectives and different types of learning for ML systems. Sherin \emph{et al.}~\cite{sherin2019systematic} also highlighted that there is no adequate empirical evidence to evaluate the effectiveness of the available testing techniques, even though the area of ML testing keeps growing rapidly. In contrast, Ricco \emph{et al.}~\cite{riccio2020testing} concluded that the most active research in ML testing has been dedicated on solving automatic test input generation and test oracle creation. Further studies are required to address numerous open challenges such as inventing proper testing metrics as well as benchmarks for ML systems.

\subsection{Interactive rapid reviews}
In this study, we use the guidelines for IRRs in software engineering proposed by Rico \emph{et al.}~\cite{Rico2020GuidelinesFC}.
Rapid reviews are a form of knowledge synthesis widely used in medicine to provide information quickly for decision-making.
As an example, during the COVID-19 pandemic, a considerable amount of rapid reviews were conducted to support decision-making in many areas of medicine~\cite{tricco2020rapid}. 
A group of researchers in software engineering proposed the use of rapid reviews to support decision-making in software projects~\cite{Cartaxo2020}.
A difference with the guidelines adopted in this study is the focus on the interaction between researchers and practitioners to make the reviews more relevant in the practitioners' context and foster the knowledge exchange. 

The guidelines are presented as a series of steps. The first step is to \textit{prepare the review}. In this step, an initial area and topic are identified, and the team is set. 
The second step is to \textit{identify review questions and prepare the IRR protocol}, where based on the exchange, the IRR team formulate review questions that represent the common interest and plan the steps to conduct the IRR i.e, IRR protocol. 
Then, the third step is to \textit{search and select papers}. In this step, shortcuts are used to reduce the space of search. For that reason, it is suggested that the review questions are narrowed to specific questions. 
Then, during the fourth step, the IRR team \textit{extracts and synthesize data} from the research literature and prepare the actions to \textit{disseminate IRR results} in the fifth step. 
It is important to clarify that the guidelines are flexible and may require adaption to the specific needs of the case. 

The software engineering research community is familiar with systematic literature reviews (SLRs) as a form of knowledge synthesis. 
Although IRRs and SLRs are similar in many aspects, like methodology and the need for a systematic approach,
it is important to clarify that IRRs do not pretend to be an alternative to SLRs when synthesising research literature. IRRs do not aim to be extensive, but provide rapid and valid input for the practitioners.
IRRs address more narrow questions than SLRs.
When conducting IRRs the review team applies shortcuts to narrow the search space and then save time in selecting and extracting relevant data. These shortcuts may result in missing relevant sources for the IRR.
There are two main reasons to select an IRR for this study. Compared with SRLs, IRRs require less resources and can be completed in shorter time frames. 
Second, IRRs, as presented here, aim to promote exchange between researchers and practitioners, which is desirable at this phase of Axis.

\subsection{Case context} \label{sec:casecontext}
Axis was the first industrial collaboration partner in AIQ. Within Axis, we identified a development team that develop solutions based on advanced ML-based computer vision. The team, develops people counting applications for dynamic environments such as shopping malls and public squares.

People counting is considered a ``statistical application'' that should be accurate on average. Corner cases are largely ignored, i.e., if a person wearing a ``funny hat'' is missed or double-counted is ok -- as long as the counter is not incremented by an amount large enough to noticeably affect the hourly/daily statistics. This is in contrast to security surveillance applications for which corner cases are critical, e.g., possible intruders crawling under the camera. Still, accuracy over time is important to people counting. False positives (counting ghosts) and false negatives (missing people) are considered equally bad. Thus the F1-score (balanced harmonic mean of precision and recall) is the primary evaluation metric. 

A set of test datasets representing scenarios in various operational environments is used for regression testing. As there are significant differences between operational environments, referred to as scenes, F1-scores are measured for individual scenes rather than for a single diverse test set. To provide reliable quality assurance, ensuring a high coverage of scenarios in the test dataset is essential. Differently sized regression test suites are running 1) in a continuous integration context, 2) on nightly builds, and 3) weekly. Two different test setups are used in the regression testing. One testing the algorithms involved only and one testing the actual hardware used.


\section{Method} \label{sec:method}
To initiate a collaboration on ML testing between researchers and practitioners at Lund University, RISE Research Institutes of Sweden, and Axis, we conducted an IRR~\cite{Rico2020GuidelinesFC} following the five steps in Table~\ref{tab:IRR-protocol}. The expected outcome of the review was threefold: 1) to establish a common view of the general problem domain -- ML testing, 2) to gain a quick overview of how current research matches with the specific needs at Axis, and 3) to propose a study aiming at filling one of the identified gaps. The researchers'  activities were carried out by the first three authors of this paper, while the fourth author and his colleagues represent the practitioner's side. Finally, the fifth author guided and monitored the research procedure.


\begin{table}[h]
\caption{Description of the five steps and corresponding research activities for this IRR study}
\label{tab:IRR-protocol}
\begin{tabular}{| p{2.3cm} | p{2cm} | p{3.3cm} |}
\toprule
\textbf{Step}                                                  & \textbf{Activities for   Researchers}                                     & \textbf{Activities for   Practitioners*}                                                                                                                         \\ \midrule
1. Prepare the review                                          & Describe research area and preliminary research goals                          & Meeting to identify mutual information needs and agree on involvement.                                                                                        \\ \midrule
2. Identify review questions and develop the IRR protocol                                 & Propose SERP taxonomy, elaborate review questions and scope of search and selection.                                           & Meeting to validate and refine taxonomy and elaborate on questions and scope. List and prioritize review questions. \\ \midrule
3. Search and select papers                                     & Map primary studies to review questions, iterate samples of selected studies with practitioners, update inclusion/exclusion criteria based on feedback.                 & Give feedback on relevance and applicability of selected papers.                                                                                                                 \\ \midrule
4. Extract and synthesize data                                 & Identify and assess   maturity of technological rules.                     & Meeting to discuss results (esp. relevance, applicability) of technological rules. Discuss how and to whom results should be summarized and communicated.     \\ \midrule
5. Disseminate IRR results. & Design and present visual abstracts for the identified TRs. Propose new research studies. & Meeting to give feedback on results. Present results within company.                                                                                                                                      \\ \bottomrule
\end{tabular}

\end{table}

The steps of an IRR are similar to the steps of other types of systematic literature reviews but adapted to meet the specific needs of an industrial stakeholder. In our case the stakeholder was Axis.

\subsection{Preparing the review}
The goal of the preparation step was to form a review team of both researchers and practitioners and to identify mutual interests and information needs with respect to the general research topic, ML testing. In this step, the interaction between industry and academia took place in an input meeting. To prepare for the input meeting, the second author put together an overview presentation of the state-of-the-art of ML testing, and the fifth author put together an overview presentation of the IRR approach. Four researchers and four practitioners took part in the meeting. The four practitioners had different roles (expert engineer, software test engineer, senior software engineer, and technical leader) at the company and thus different perspectives on the topic. At the meeting, after the presentations, the practitioners shared aspects of their practices and challenges of testing their ML applications. 

\subsection{Identifying review questions and developing the review protocol}
The goal of the second step was to agree on a list of prioritized review questions and an initial review protocol. Here interactions took place in a workshop and a follow-up ranking exercise. Before the workshop, the first three authors developed a preliminary SERP-taxonomy~\cite{petersenSERP_2014} based on the state-of-the-art of ML testing and previous SERP-taxonomies on software testing~\cite{engstrom_SERP-test_2017,Ali_search_2019}. A SERP-taxonomy includes four facets (i.e., scope, context, effect, and intervention) to align descriptions of research solutions and industry challenges. In that way, such a taxonomy may be used to facilitate communication between practitioners and support the mapping of challenges from the industry to available research~\cite{engstrom_SERP-test_2017}. 

During the workshop, including the full review team, we walked through all facets and entities of the taxonomy to trigger discussions about ML testing challenges and potential solutions from various perspectives. Based on the outcome of the workshop, the researchers updated the taxonomy and proposed a list of 12 potentially relevant review questions to Axis. This list was then sent out to all participants (researchers and practitioners) with a request to rank them in order of interest using an ordinal scale from 1 to 5. After summarizing the results of this exercise, we agreed to search for research relevant to the three highest ranked questions. Furthermore, we agreed to include research based on relevance and applicability for Axis, but did not specify this further at this point. 

\subsection{Searching and selecting primary studies}
During this step, we successively refined the review protocol while searching for relevant studies and delimiting the scope (i.e., defined exclusion criteria). As part of this activity, the researchers conducted the search and selection while the practitioners gave feedback on relevance and applicability of a small sample of papers sent to them by email. 

We limited the search to the research covered by three recent secondary studies on ML testing~\cite{zhang2020machine, riccio2020testing, sherin2019systematic}. The researchers revisited the complete set of primary studies (180 papers) to map them to the three review questions. This screening was based on full-text scanning as it was impossible to do it based neither on the original classification (in the secondary studies) nor on a sole title and abstract screening. At this stage, we had an inclusive approach meaning that if any of the three researchers marked a paper as relevant for a review question, it was coded as such. 

Due to the large number of papers coded as potentially relevant to at least one of the three review questions, we decided to descope further and focus the in-depth analysis only on one of the review questions. Thus, the continued selection focused only on the highest prioritized review question, i.e., ``How to test the dataset?'' 35 of the 180 studies were marked as potentially relevant for this question. After a thorough review, only five of them remained. At this stage, the remaining candidates were tightly connected to the Axis' context, i.e., testing data for ML-based computer vision. 

\subsection{Data extraction and synthesis}
From the selected papers, we extracted technological rules following the design science lens described by Storey \emph{et al.}~\cite{storey_2017_using}. A technological rule is a structured way of describing research contributions with respect to their effect, context, and intervention. Technological rules can be extracted and presented at different levels of abstraction, and are used for communicating the research output in a simple and condensed way~\cite{storey_2017_using}.
Our goal was to compare technological rules, identify research gaps and the specific needs at Axis. To allow for generalization, we extracted technological rules of different abstraction levels from the primary studies. Furthermore, we extracted the maturity of the rules in terms of empirical observations and analytical reasoning that supported the propositions. The technological rules were then presented to the review team in a short reaction meeting, where the industrial team provided their reflections about relevance and applicability at Axis. 

\subsection{Disseminating the review results}
As the main goal of the review was to initiate a new collaboration, we did not have a plan for disseminating the results within Axis. Instead the most relevant technological rules provide input for new MSc thesis proposals. However, unplanned dissemination took place as new knowledge travel between teams. The technological rules from one of the included papers were evaluated for another purpose by another team at Axis. On the academic side, results were summarized and presented as visual abstracts, one for each technological rule, at a seminar and in this report. 

\subsection{Validity of contributions}
The main goal of conducting an IRR is not to build general theory or publish rigorous research results, but to extract sufficient knowledge to act in a specific situation. In our case, our primary goal was to align terminology, match interests, and initiate industry-academia collaboration on ML testing with one case company. Still, we believe the reported contributions could be helpful for other researchers with similar goals but they must be adapted to their contexts.

The taxonomy builds on previous work on taxonomy development~\cite{petersenSERP_2014}, general testing terminology~\cite{engstrom_SERP-test_2017} and recent ML testing syntheses~\cite{sherin2019systematic, riccio2020testing,zhang2020machine}. Hence, they are well founded in the research literature. However, the industrial validation is made from a narrow perspective in a single case context. Thus, when elaborating the taxonomy, the details mirror the review team's interests and experiences, including four practitioners and four researchers.

Similarly, the list of challenges represents a single case, and the ranking of importance represents the interests of the review team. At a high abstraction level, these challenges and interests match the interests of the research community, represented by the 180 primary studies.

A review may also be validated based on its coverage, i.e., is any important work missing? In our case, we did not conduct an extensive search on our own but relied on the rigorous searches made in three recent secondary studies on the same topic. Although their searches were extensive, the time delay caused by the publication process leads to the omission of the most recent publications, i.e., from 2020 and onward. Thus, more recent research may provide a better match to our review questions. This should be considered in future work. Nevertheless, conclusions regarding relevance and applicability of available research are still valid and may guide future reviews.

\section{Results} \label{sec:res}

In this section we present the outcome of each step of the IRR. Subsection~\ref{sec:result_taxonomy} describes the entities of the taxonomy after validation,~\autoref{sec:result_open_questions} lists the open questions derived from the workshop,~\autoref{sec:result_mapping} describes our mapping of primary studies to the review questions,~\autoref{sec:exclusion_criteria} presents the final exclusion criteria resulting from the iterative review of papers mapped to review question~1,~\autoref{sec:result_data_testing} presents the nine technological rules extracted from the five most relevant primary studies, and~\autoref{sec:result_gap} finally elaborates on what we did not find in the research literature, i.e., the gap between research and practice in our case. 
\subsection{ML testing taxonomy}
\label{sec:result_taxonomy}
As a result of the initial interaction between the case company and the researchers, we agreed on creating a taxonomy of ML testing to guide the collaboration further. We developed a general taxonomy for three out of four facets in the SERP-taxonomy architecture~\cite{petersenSERP_2014} (i.e., context, scope, and effect). Since we enter this review from the challenge perspective, we found that detailing these three facets were sufficient for the communication within the review team. The fourth SERP-facet, intervention, may be used to elaborate technical aspects that support abstraction and comparison of classes of solutions.

Our resulting taxonomy, presented in Figure~\ref{fig:taxop1}, aims to guide the formulation of practical ML testing challenges at an appropriate abstraction level to support identification and design of relevant research.

\subsubsection{Scope of ML testing interventions}
The green sector in Figure~\ref{fig:taxop1} shows the details of the scope facet. Scope here refers to the testing activity, or part of the testing process, on which a potential intervention may focus. We identified four important aspects where two are derived from generic testing literature, i.e., testing process and testing levels, and the other two are specific to the ML context, i.e., parts of the ML-system to be tested and the mode of operations (online or offline testing, cf. Figure~\ref{fig:taxop1}).


\subsubsection{Effect of ML testing interventions}
The effect of an intervention is described in terms of its observed or desired impact on the testing. It could for example be the reported result of an empirical evaluation or an identified practical need in an exploratory study or case description. The blue sector in Figure~\ref{fig:taxop1} shows the effect facet. The main types of desired effects are derived from the generic SERP-taxonomy architecture~\cite{petersenSERP_2014}, i.e., solving an unsolved problem (solve), improving the current solution (improve), and assessing the current situation (diagnose). 


\subsubsection{Context of ML testing interventions}
The context facet aims to capture factors in the context that impact the effect or applicability of an intervention. Several such factors were identified, reflecting the multidimensional variation of both ML systems and testing approaches. Eight aspects of the context were included as shown in the yellow sector of Figure~\ref{fig:taxop1}: 1) programming languages used for the implementation of the system, 2) degree of access to the system components, 3) framework, 4) machine learning type, 5) testing setup, 6) application, 7) type and 8) domain of the system. 

At the initial stages of our project, we used the taxonomy for structuring the information in the secondary studies and for triggering discussions about the topic within the review team. We believe the resulting taxonomy may be used in similar ways by others to view ML testing challenges and solutions from various perspectives and thus support communication between researchers and practitioners who approach the topic -- in different ways and in many cases at different abstraction levels. While conducting this review, we experienced that considering all the facets and digging in to details of all the facets of the taxonomy helped the communication and, by extension, our common understanding.


\FloatBarrier

\subsection{Prioritized list of open questions}
\label{sec:result_open_questions}
\begin{figure*}[h!]
\centering
\includegraphics[width=\textwidth]{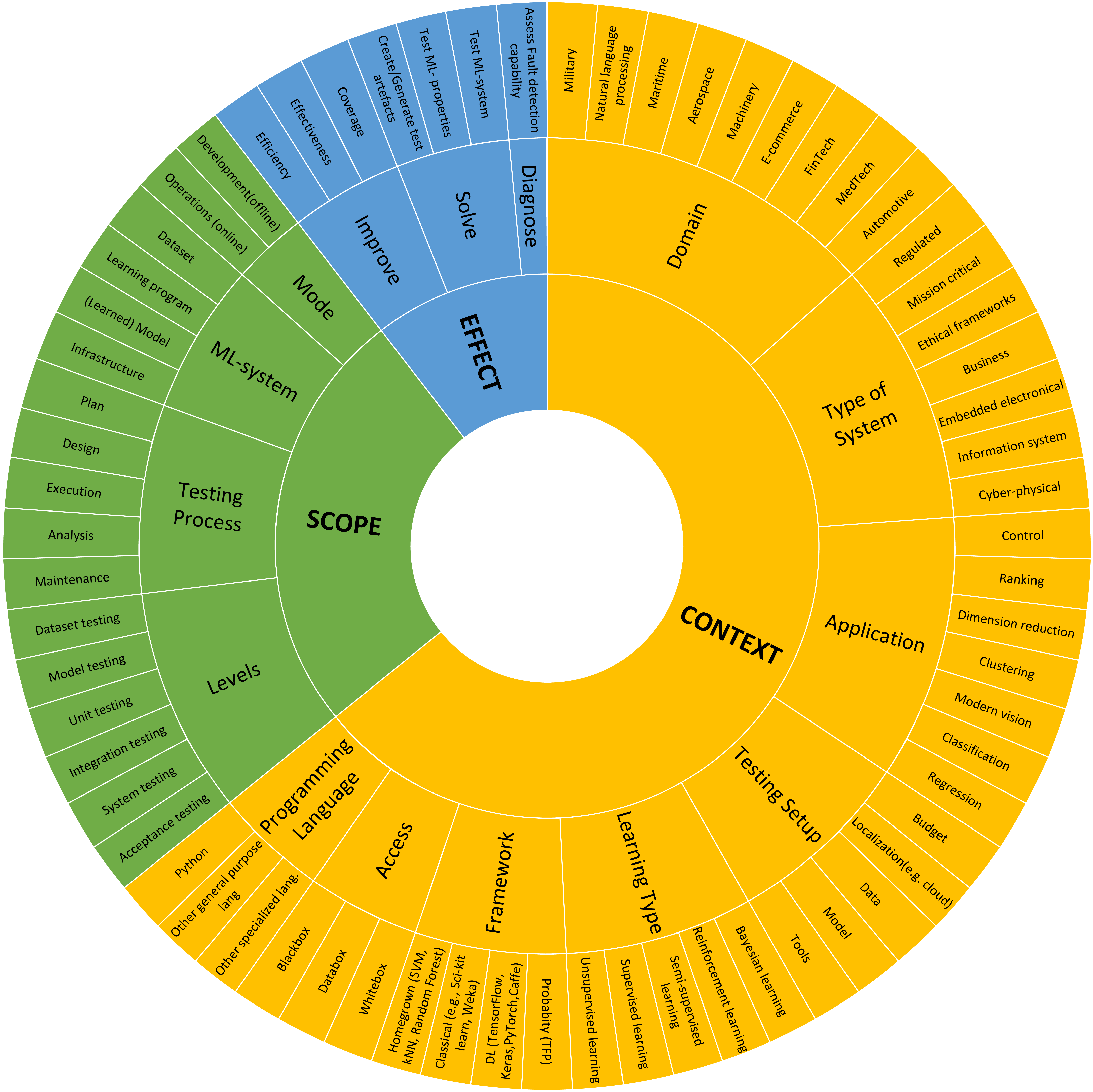}
\caption{An ML testing SERP taxonomy with the facets context, scope, and effect.}
\label{fig:taxop1}
\end{figure*}

Guided by the taxonomy and the feedback from practitioners in the initial meetings, we formulated 12 review questions potentially relevant to our case. In the following list, organized using the scope facet of the taxonomy in Figure~\ref{fig:taxop1}, boxes represent the review team's highest priority questions. Verbs from the effect facet appear in italics.
\begin{itemize}
\item \textbf{ML system -- Dataset}	
\begin{enumerate}
\begin{framed}
    \item[(\#1)] How to test the dataset? The dataset is evolving, which motivates the following sub-questions: 
    \begin{enumerate}
        \item How to identify mislabeled data in the dataset?
        \item Adequacy testing, i.e., how to assess and improve data (scenario) coverage of the training and test datasets in terms of diversity?
        \item How to assess potential bias after the training/test split?
        \item How valid is the data and its use? Is the data used for testing within the operational design domain? Or did some of the data originate from another source?
  \end{enumerate}
      \end{framed}
\end{enumerate}
\item \textbf{ML System -- Learning program}	
\begin{enumerate}[resume]
    \item[(\#2)] How to \textit{diagnose} (assess the fault detection capability) and \textit{improve} (unit) testing (design and analysis) of the learning program?
    \item[(\#3)] How to \textit{improve} testing of the learning program to detect more faults? (e.g., using unit testing)
\end{enumerate}
\item \textbf{ML system -- Learned model}
\begin{enumerate}[resume]
\begin{framed}
\item[(\#4)] Are there complementary metrics to assess model correctness (accuracy)? (e.g., edge case measures, uncertainty scores, aggregated metrics across scenes)\end{framed}
\item[(\#5)] How to interpret and analyze the testing result of the ML model? (e.g., increased automation or visual analytics)
\item[(\#6)] How to \textit{diagnose} whether the current set of scenarios in the test dataset are appropriate for detecting faults on the model level?
\item[(\#7)] How to \textit{improve} coverage testing with respect to scenario diversity?
\item[(\#8)] How to \textit{improve} the test dataset to increase fault detection ability?
\item[(\#9)] How to \textit{improve} test prioritization to increase regression test efficiency?
\begin{framed}
\item[(\#10)] How to generate new test cases for testing the model? (e.g, synthetic data, data augmentation, guided search)\end{framed}
\end{enumerate}
\item \textbf{Levels -- System testing} (see Section~\ref{sec:casecontext})	
\begin{enumerate}[resume]
\item[(\#11)] How to \textit{improve} (acceptance) testing of the ML-based system? 
\item[(\#12)] How to \textit{diagnose} whether the current set of scenarios in the test dataset are appropriate for detecting faults on the system level?
\end{enumerate}
\end{itemize}
Even though the above questions represent the interests of the review team, we argue that they constitute real challenges that generally deserve attention. We believe they may support other researchers trying to identify research gaps in ML testing for computer vision applications.

\subsection{Mapping primary studies to the three most important open questions}
\label{sec:result_mapping}
The review team (i.e., four practitioners and three researchers) ranked the questions based on importance. Then the 180 primary studies covered in this review were mapped to the three review questions that were considered the most important (i.e., number \#1, \#4, and \#10 (in the list of questions above). 35 primary studies were marked as potentially relevant for review question 1, 25 primary studies for review question 4, and 68 for review question 10. The details of this mapping can be found on Zenodo~\cite{qunying_song_2022_5865070}. 

This mapping may help navigating the research literature and could be used as a starting point by researchers and practitioners facing similar challenges.

\subsection{Selection of studies related to question \#1}
\label{sec:exclusion_criteria}
This section describes our analysis of relevance and applicability for Axis in relation to the review question \#1. The final list of exclusion criteria is:
\begin{itemize}
    \item \emph{Purpose}. The proposed mechanism does not evaluate some properties of the data.
In MLware, the training dataset is part of the system~\cite{borg_AIQ_2021}. The purpose of the proposed mechanism shall be to test these data, not to generate synthetic data -- unless the synthetic data are specifically used to validate the training data through comparison. 
Examples of excluded papers relate to: 
\begin{itemize} 

\item Test data generation for ML systems such as Zhang \emph{et al.}~\cite{zhang2018deeproad} and Tian \emph{et al.}~\cite{tian2018deeptest} that generate artificial driving scenes for testing ML-based autonomous driving functions.
\item Testing the model fitness like Zhang \emph{et al.}~\cite{zhang2019perturbed} which validates the model relevance, and ML model underfitting as well as overfitting using a perturbed model validation technique.
\item Online monitoring of data prior to making predictions, e.g., Henriksson \emph{et al.}~\cite{henriksson2019towards}, since it targets testing aspects of the operational environment and detects inputs that are outside the training dataset.
\item The proposed mechanism is intended to protect the neural network from antagonistic attacks. Antagonistic attacks are closer to cybersecurity research than data validation. Furthermore, adversarial attacks would typically target the system in operation and not the training/validation/test dataset. An example of such studies is Uesato \emph{et al.}~\cite{uesato2018rigorous} which evaluates learning systems in safety-critical domains by identifying the adversarial situations.
\end{itemize}
\item \emph{Applicability}. The proposed mechanism is not applicable in the Axis' context. Reasons for exclusion include:
\begin{itemize}
\item \emph{Not NN learning}.
The mechanism is not applicable to supervised learning with neural networks. Axis uses neural networks for supervised learning and the proposed mechanism must be applicable. Thus, papers that explicitly address other learning mechanisms are considered out of scope, such as Krishnan \emph{et al.}~\cite{krishnan2016activeclean} on support vector machines and Uesato \emph{et al.}~\cite{uesato2018rigorous} on reinforcement learning.
\item \emph{Not images}.
There is no explicit mention of how the proposed mechanism could work for images. Axis trains models for video sequences and the proposed mechanism must be applicable. Thus, interventions that validate only non-image data are excluded, e.g., spell or format checking, and named entity recognition on tabular data~\cite{hynes2017data, breck2019data, schelter2018automating}.
\end{itemize}
\end{itemize}

\subsection{Best matches}
\label{sec:result_data_testing}
After applying the criteria described above, five primary studies remained, partly answering the general review question (\#1 in Section~\ref{sec:result_open_questions} ) ``How to test the dataset?'' Although none of the proposed solutions were directly applicable in the case context, Axis confirmed related problem formulations and shared potentially valuable ideas.  
In line with the design science lens~\cite{storey_2017_using}, we extracted technological rules for each paper.
A technological rule captures the mapping between a problem and a solution. We describe them in terms of: \textit{\textbf{To achieve <effect> IN <context> DO <intervention>}}. We describe the technological rules at different abstraction levels, i.e, some are more concrete and others are more general. 


\subsubsection{Paper I}

Ma \emph{et al.}~\cite{ma2018deepmutation} propose a mutation testing framework for deep learning systems to assess the quality of the test data. Specifically, a set of source-level mutation operators are defined to introduce faults to the training data and the training programs. In addition, a set of model-level mutation operators are defined to create mutants for the deep learning models without a training process. Eventually, the effectiveness of the test data can be evaluated from the analysis based on to what extent the injected faults could be detected. The authors have used the framework on two publicly available datasets (i.e., MNIST and CIFAR-10) with three popular deep learning models, and demonstrated the effectiveness of the framework for designing and constructing high-quality test datasets for deep learning MLware.

Two technological rules were extracted from this paper~\cite{ma2018deepmutation}, i.e., a concrete one:
\begin{tcolorbox}[colback=black!5!white,colframe=black!75!white,title=Technological rule 1]
  To measure the quality of test data for deep learning systems, use an adapted form of mutation testing.
\end{tcolorbox}
and a general one:
\begin{tcolorbox}[colback=black!5!white,colframe=black!75!white,title=Technological rule 2]
  To improve the generality and robustness of deep learning models, test the test dataset.
\end{tcolorbox}

Both the source-level and model-level mutant operators are general and can be reused. Furthermore, the concept of the proposed framework, i.e., to use mutation testing, is not constrained by any specific type of deep learning application or data used. Thus, we consider this paper potentially relevant, and the synthesized findings can be transformed for testing the quality of test data for Axis as well. The response from Axis was positive, they found the approach interesting but questioned the scalability. Axis works with orders of magnitude larger datasets (about $10^5$ to $10^7$ images, which is 10 to 1,000 times larger than the MNIST and CIFAR-10 datasets) than the ones used for evaluating the approach in the paper. In addition, the industrial datasets are rather Full HD resolution than the 32$\times$32 pixels targeted in many research papers. Another question is which mutant operators would work for the complexity of the industrial case, as MNIST and CIFAR-10 are trivial datasets in comparison. A pre-study in the Axis context will be initiated to investigate this further. 

\subsubsection{Paper II}
\label{sec:surprice_adequacy}
Kim \emph{et al.}~\cite{kim2019guiding} propose a concept of surprise adequacy as the test adequacy criterion for deep learning systems. Based on the trace of the neuron activation when executing the deep learning model on both the training data and the testing data, surprise adequacy can be calculated using either a likelihood-based approach or a distance-based approach. The resulting surprise adequacy indicates how surprising the test data is compared to the training data. The concept assumes that a good test input should be sufficiently, but not excessively, surprising to the training data. The authors also evaluate the effectiveness of using the surprise adequacy metric for sampling test input and improving the model accuracy via retraining, based on publicly available datasets such as MNIST and CIFAR-10, and deep learning systems for autonomous vehicles like Dave-2 and Chauffeur.

Two technological rules were extracted from this paper~\cite{kim2019guiding}, the concrete one is:
     
\begin{tcolorbox}[colback=black!5!white,colframe=black!75!white,title=Technological rule 3]
  To improve the classification accuracy of deep learning systems, retrain the model, systematically sampling inputs based on the surprise adequacy criterion.
\end{tcolorbox}

while the general one is described as: 
\begin{tcolorbox}[colback=black!5!white,colframe=black!75!white,title=Technological rule 4]
  To test the correctness and robustness of deep learning systems, test the systems' behaviour with respect to their training data.
\end{tcolorbox}

The proposed criterion -- surprise adequacy -- and the corresponding ways of computing such a criterion are transferable to different deep learning MLware as the training data, testing data, and neuron activations are inherent parts of such systems. The expected outcome is an indication of how good or how different the test is compared to the training data.

While the surprise adequacy metric was not considered relevant for the practitioners in the review team, it was explored by another team at Axis. Here it was explored for a slightly different purpose, i.e., to guide complementary data collection rather than for testing data. Collecting additional data that strives to maximize the surprise adequacy has also been proposed by the original inventors~\cite{kim2020reducing}, as an approach to increase the diversity of both training and test datasets. However, Axis compared surprise adequacy to a set of other metrics and in the end they selected another option (recent work proposed by Pleiss et al.~\cite{NEURIPS2020_c6102b37}, not covered by the secondary studies used for our study selection). Part of the reason was that the proposed surprise adequacy calculations did not scale to size of the data set as described earlier (see subsection 4.5.1). On the other hand, Axis encourages additional research into surprise adequacy calculation for representative subsets of the data, i.e., aggregating measures for families of neuron activation traces.

\subsubsection{Paper III}
Byun \emph{et al.}~\cite{byun2019input} introduce three different metrics for test input prioritization for deep neural networks, i.e., (1) confidence, measured by using the softmax function, (2) uncertainty, measured using Bayesian Networks, and (3) surprise as described in the previous paper by Kim \emph{et al.}~\cite{kim2019guiding}. The authors apply these metrics to prioritize test inputs for two different systems (i.e., a digital classification system trained on the MNIST dataset, and TaxiNet) for image classification. They show the effectiveness of the metrics in indicating fault-revealing inputs and, by extension, for selecting test input and improving the model via retraining.  

Two technological rules were extracted from this paper~\cite{byun2019input}, including a concrete one: 
\begin{tcolorbox}[colback=black!5!white,colframe=black!75!white,title=Technological rule 5]
  To prioritize test input for deep neural networks, apply metrics of confidence, uncertainty, and surprise.
\end{tcolorbox}
and a general one: 
\begin{tcolorbox}[colback=black!5!white,colframe=black!75!white,title=Technological rule 6]
  To increase test effectiveness when testing deep neural networks in safety-critical systems, prioritize test input.
\end{tcolorbox}

Similar to paper~II, this paper proposes metrics to measure the sentiment of the deep neural networks. Then, the test inputs can be validated, prioritized, and effectively selected for testing and retraining the model. The findings are considered generic and potentially relevant for Axis' needs. While ``Surprise Adequacy'' was investigated by another team at the company (see subsection~\ref{sec:surprice_adequacy}) the other two metrics ``confidence'' and ``uncertainty'' have not yet been considered.

\subsubsection{Paper IV}

Cheng \emph{et al.}~\cite{cheng2018towards} study a set of metrics to measure the dependability attributes of neural networks.
The metrics include robustness, interpretability, completeness, and correctness. In our review, the paper was initially excluded due to its purpose (i.e., measuring the dependability of neural network models) and the application on autonomous driving. However, after a second review, the paper was included since the part related to the completeness of the training data seems relevant. The paper uses neuron k-activation and neuron activation patterns as a measure of scenario coverage and completeness of training data for NN-based autonomous driving systems. The assessment of completeness of the training data, which is used for testing the coverage of the training data but is general for testing of data regardless of its use, could support both quality assurance of training or test datasets.

One general technological rule was extracted from this paper~\cite{cheng2018towards} as only the part about the training data completeness is relevant. The technological rule is described as:
\begin{tcolorbox}[colback=black!5!white,colframe=black!75!white,title=Technological rule 7]
  To measure dependability of neural networks, evaluate the completeness of the training data.
\end{tcolorbox}

The part that involves measuring completeness of training data is relevant for data testing in general. While the paper sets the general focus in autonomous driving applications, further investigation on how it could be applied into Axis' context should be performed. 

\subsubsection{Paper V}
Bolte \emph{et al.}~\cite{bolte2019towards} construct a system framework for corner case detection in training datasets for autonomous driving systems. The framework consists of three parts: (1) a semantic segmentation model to partition and classify the image into different semantic parts; (2) an image prediction model that predicts the next image based on the previous set of images and counting the errors based on the real image; and (3) a detection system that detects the corner cases if an object is unpredictable given the error score counted in the previous model. The proposed framework can be used in both online and offline modes. The difference is that the offline mode takes a collected database of image data for training, and the online model uses live video frames collected by the camera installed on vehicles. The authors have trained and evaluated the framework on the Cityscapes dataset and achieved prominent results for detecting unusual situations for autonomous driving. 
Two technological rules were extracted from this paper~\cite{bolte2019towards}, the concrete one is: 
\begin{tcolorbox}[colback=black!5!white,colframe=black!75!white,title=Technological rule 8]
  To detect corner cases for ML-based autonomous driving systems, use a system framework based on image segmentation and prediction.
\end{tcolorbox}
while the general one is:
\begin{tcolorbox}[colback=black!5!white,colframe=black!75!white,title=Technological rule 9]
  To improve the robustness of machine learning systems, identify critical situations.
\end{tcolorbox}

The concept of the study is relevant for data testing for ML systems in general, although the proposed system framework works with image data. It predicts corner cases for autonomous driving, which can be used to test and retrain the ML model. Still, the actual applicability of using this framework in Axis needs to be further investigated.

While identifying corner cases is not so important for the application of people counting, which is the focus of the practitioners in the review team, it could be vital for the companies' other products in the security business segment. For example, a corner case of someone moving strangely to avoid detection would be critically important for a security application to detect. For security applications, detecting such anomalies are among the most important use cases. On the other hand, for people counting applications, it could be interesting to apply a broader definition of corner cases, i.e., not only very rare cases but rather underrepresented scenarios. Any mechanism that could support identification of such scenarios would be helpful for ML testing.

\subsection{Identified gap}
\label{sec:result_gap}
The five papers that we have presented in the previous subsection mainly include frameworks, metrics, constructed tools, and mechanisms for measuring the quality of the data and consistency of the test data with respect to the training data for ML systems. Based on the general focus and sub-questions of the first review question (i.e., how to test the dataset), we observed that the extracted technological rules can be used to address some, but not all, related perspectives. 

In particular, there are no studies that provide relevant techniques for a) identifying mislabeled data in the dataset. Thus, this sub-question is still an open challenge and needs to be studied further. However, paper IV provides mechanisms for measuring the completeness of training data, which gives some insights and potential solutions for b) assessing and improving the training and testing data coverage -- also from the perspective of scenario coverage.
Note that the proposed mechanism focuses on the autonomous driving domain. In addition, no studies we identified target c) assessing biases between the training and testing data. 
However, papers~II and III could be potentially relevant since they support measuring and filtering test data not represented in the training data using different metrics (i.e., confidence, uncertainty, and surprise adequacy). The same observation also applies for sub-question d) how valid the data used for testing is with respect to the training data, where the metrics proposed in papers~II and III can be used for such purposes. As a result, we believe the findings we synthesized can ease some parts of the research question we focus on, whereas further investigation is needed to address the remaining gaps.



\section{Discussion} \label{sec:disc}
AI engineering is an emerging field that is vital for AI quality. As argued in Section~\ref{sec:bg}, ML testing is going to play a critical role in ensuring that future AI solutions are trustworthy. However, there is no established go-to model describing ML testing. Several previous studies propose dimensions to bring structure to the research area. This paper synthesizes a novel taxonomy based on three secondary studies. The taxonomy shall be considered work-in-progress, but it already has provided value for us in an emerging industry-academia collaboration.

The three secondary studies used three different classification strategies for their respective goals. Both Riccio \textit{et al.}  and Sherin \textit{et al.} refer to their works as systematic mapping studies and focus on trends and gaps. The scope of Riccio \textit{et al.} is functional testing of ML systems and they structure the 70 primary studies based on 1) system context, 2) testing approach, and 3) empirical evaluation~\cite{riccio2020testing}. Sherin \textit{et al.}'s mapping, including also non-functional testing, provides less synthesis and rather extracts fine-granular information from the 37 primary studies. In the secondary study by Zhang \textit{et al.}, referred to only as a survey, the authors explicitly specify their ambition to provide a comprehensive overview of ML testing. The 138 included papers are organized into 1) testing properties, 2) testing components, 3) testing workflow, and 4) application scenarios. We believe that our proposed taxonomy combines the complementary perspectives provided in previous work.


Any taxonomy or model is developed for a specific communication purpose, targeting a defined group of people. In our case, the goal was to align terminology and identify shared interests within the group of researchers and practitioners. Thus, we found the SERP approach~\cite{petersenSERP_2014} applicable and useful. 
Furthermore, by building on SERP-test~\cite{engstrom_SERP-test_2017}, comprising common testing terminology, and adding the ML perspective from the secondary studies~\cite{sherin2019systematic, riccio2020testing, zhang2020machine} as well as from the case company, we got a solid basis for our communication on the topic -- ML testing.


At a general level, we found a good match between the prioritized challenges of our case company and the research focus within the community. As a result, 92 out of 180 primary studies were classified as potentially relevant for at least one of the top three review questions, i.e., \#1 How to test the dataset (35 primary studies), \#2 How to assess model accuracy (metrics) (25 primary studies), and \#3 How to generate test cases for testing the model (68 primary studies). However, as shown in the in-depth analysis of the first set of papers, no perfect matches exist.


Of the 35 primary studies initially considered relevant for data testing, we finally selected and reviewed five that best matched the specific context and needs at Axis. After analyzing and synthesizing the findings from the papers, we extracted nine technological rules. The practitioners were positive to the presented techniques (see subsection 4.5) and thought most of them could be relevant. Particularly, they have used the surprise adequacy for complementing data collection in the company as described in subsection 4.5.2. However, we found no perfect match directly transferable to the applications and data testing issues at Axis. As underlined by the definition of AI quality~\cite{borg_AIQ_2021}, finding feasible ways to perform data quality assurance is at the heart of the problem. Therefore, we are convinced that data testing will play an important role in the future of AI engineering. Also, it is significant in future research to explore further how to instantiate and evaluate the techniques in this industrial setting. 


The concepts (e.g., metrics and criteria) and interventions (e.g., approaches and frameworks) presented in the five papers are quite generic for data testing in the ML field as to the extent of our interpretation. Hence, we believe those concepts and interventions can be reused and adapted for solving potential issues for different ML application domains. In the same way, the technological rules can be used to map solutions to challenges at different abstract levels, and support the communication and knowledge exchange between academic researchers and industrial practitioners in further studies.






\section{Conclusions} \label{sec:conc}
We report outcomes and lessons learned from applying an interactive rapid review on machine learning testing. The review team consisted of four researchers from Lund University and RISE Research Institutes of Sweden and four practitioners from Axis. The primary goal of the study was to initiate collaboration and align terminology and interests between the partners. 

Three secondary studies, covering 180 primary studies on machine learning testing, functioned as a starting point for the review. The classifications of research in the secondary studies were mapped to the SERP taxonomy architecture~\cite{petersenSERP_2014} to guide the alignment of terminology and interests within the review team. The resulting SERP taxonomy were further extended by general taxonomies on software testing~\cite{engstrom_SERP-test_2017} built on the same taxonomy architecture. Finally, we validated and updated the outcome based on discussions and reflections in the review team. While we plan to evolve the outcome, \emph{this paper presents the latest version of the taxonomy.}

The new SERP taxonomy was used to identify and describe current challenges in the case context. \emph{The complete list of challenges are presented in this report}. Furthermore, the review team ranked the challenges by their perceived importance for the target organizational unit within the case company.

The primary studies were mapped to the three most important questions. Moreover, we conducted an in-depth analysis of the 35 papers for the highest ranked question, i.e., ``How to test the dataset?'' \emph{We present and discuss 9 technological rules on data testing}, extracted from 5 of the papers. \emph{Finally, we report and discuss the relevance and applicability criteria used to filter out those 5 papers.} 

As AI quality combines source code and data quality~\cite{borg_AIQ_2021}, we believe that data testing will be increasingly important within the field of AI engineering. Our findings call for more research on the topic, not the least for image data, targeting business-critical computer vision systems. Furthermore, convincing data testing for computer vision applications can potentially constitute a cornerstone in the safety argumentation in future assurance cases, e.g., for critical ML-based perception applications in automotive~\cite{borg2019safely}, avionics~\cite{vidot2021certification}, and healthcare~\cite{jiang2017artificial}.  

The motivation for this interactive rapid review was to identify research or research gaps of relevance for the case company. Thus, all steps in the process have been guided by Axis' specific needs. As our next step, we plan to design a joint solution-oriented study on the topic of data testing as well as a set of MSc thesis project proposals. Based on the discussions of the selected studied in Section~\ref{sec:result_data_testing}, there are several promising directions for future collaborations. Our case (of industry-academia collaboration) is a single case and as such a proof-of-concept that may be extended with additional cases. 

\section*{Acknowledgments}
This initiative received financial support through the AIQ Meta-Testbed project funded by Kompetensfonden at Campus Helsingborg, Lund University, Sweden. In addition, this work was supported in part by the Wallenberg AI, Autonomous Systems and Software Program (WASP). 

%% file: data_testing.bbl

\begin{thebibliography}{47}


\ifx \showCODEN    \undefined \def \showCODEN     #1{\unskip}     \fi
\ifx \showDOI      \undefined \def \showDOI       #1{#1}\fi
\ifx \showISBNx    \undefined \def \showISBNx     #1{\unskip}     \fi
\ifx \showISBNxiii \undefined \def \showISBNxiii  #1{\unskip}     \fi
\ifx \showISSN     \undefined \def \showISSN      #1{\unskip}     \fi
\ifx \showLCCN     \undefined \def \showLCCN      #1{\unskip}     \fi
\ifx \shownote     \undefined \def \shownote      #1{#1}          \fi
\ifx \showarticletitle \undefined \def \showarticletitle #1{#1}   \fi
\ifx \showURL      \undefined \def \showURL       {\relax}        \fi
\providecommand\bibfield[2]{#2}
\providecommand\bibinfo[2]{#2}
\providecommand\natexlab[1]{#1}
\providecommand\showeprint[2][]{arXiv:#2}

\bibitem[\protect\citeauthoryear{Ali, Engstr{\"o}m, Taromirad, Mousavi, Minhas,
  Helgesson, Kunze, and Varshoaz}{Ali et~al\mbox{.}}{2019}]%
        {Ali_search_2019}
\bibfield{author}{\bibinfo{person}{{Nauman Bin} Ali}, \bibinfo{person}{Emelie
  Engstr{\"o}m}, \bibinfo{person}{Masoumeh Taromirad},
  \bibinfo{person}{Mohammad Mousavi}, \bibinfo{person}{{Nasir Mehmood} Minhas},
  \bibinfo{person}{Daniel Helgesson}, \bibinfo{person}{Sebastian Kunze}, {and}
  \bibinfo{person}{Mahsa Varshoaz}.} \bibinfo{year}{2019}\natexlab{}.
\newblock \showarticletitle{On the search for industry-relevant regression
  testing research}.
\newblock \bibinfo{journal}{\emph{Empirical Software Engineering}}
  \bibinfo{volume}{24}, \bibinfo{number}{4} (\bibinfo{year}{2019}),
  \bibinfo{pages}{2020--2055}.
\newblock
\showISSN{1573-7616}
\urldef\tempurl%
\url{https://doi.org/10.1007/s10664-018-9670-1}
\showDOI{\tempurl}


\bibitem[\protect\citeauthoryear{Amershi, Begel, Bird, DeLine, Gall, Kamar,
  Nagappan, Nushi, and Zimmermann}{Amershi et~al\mbox{.}}{2019}]%
        {8804457}
\bibfield{author}{\bibinfo{person}{Saleema Amershi}, \bibinfo{person}{Andrew
  Begel}, \bibinfo{person}{Christian Bird}, \bibinfo{person}{Robert DeLine},
  \bibinfo{person}{Harald Gall}, \bibinfo{person}{Ece Kamar},
  \bibinfo{person}{Nachiappan Nagappan}, \bibinfo{person}{Besmira Nushi}, {and}
  \bibinfo{person}{Thomas Zimmermann}.} \bibinfo{year}{2019}\natexlab{}.
\newblock \showarticletitle{Software Engineering for Machine Learning: A Case
  Study}. In \bibinfo{booktitle}{\emph{2019 IEEE/ACM 41st International
  Conference on Software Engineering: Software Engineering in Practice
  (ICSE-SEIP)}}. \bibinfo{pages}{291--300}.
\newblock
\urldef\tempurl%
\url{https://doi.org/10.1109/ICSE-SEIP.2019.00042}
\showDOI{\tempurl}


\bibitem[\protect\citeauthoryear{Bjarnason, Runeson, Borg, Unterkalmsteiner,
  Engstr{\"o}m, Regnell, Sabaliauskaite, Loconsole, Gorschek, and
  Feldt}{Bjarnason et~al\mbox{.}}{2014}]%
        {bjarnason2014challenges}
\bibfield{author}{\bibinfo{person}{Elizabeth Bjarnason}, \bibinfo{person}{Per
  Runeson}, \bibinfo{person}{Markus Borg}, \bibinfo{person}{Michael
  Unterkalmsteiner}, \bibinfo{person}{Emelie Engstr{\"o}m},
  \bibinfo{person}{Bj{\"o}rn Regnell}, \bibinfo{person}{Giedre Sabaliauskaite},
  \bibinfo{person}{Annabella Loconsole}, \bibinfo{person}{Tony Gorschek}, {and}
  \bibinfo{person}{Robert Feldt}.} \bibinfo{year}{2014}\natexlab{}.
\newblock \showarticletitle{Challenges and practices in aligning requirements
  with verification and validation: a case study of six companies}.
\newblock \bibinfo{journal}{\emph{Empirical software engineering}}
  \bibinfo{volume}{19}, \bibinfo{number}{6} (\bibinfo{year}{2014}),
  \bibinfo{pages}{1809--1855}.
\newblock


\bibitem[\protect\citeauthoryear{Bolte, Bar, Lipinski, and Fingscheidt}{Bolte
  et~al\mbox{.}}{2019}]%
        {bolte2019towards}
\bibfield{author}{\bibinfo{person}{Jan-Aike Bolte}, \bibinfo{person}{Andreas
  Bar}, \bibinfo{person}{Daniel Lipinski}, {and} \bibinfo{person}{Tim
  Fingscheidt}.} \bibinfo{year}{2019}\natexlab{}.
\newblock \showarticletitle{Towards corner case detection for autonomous
  driving}. In \bibinfo{booktitle}{\emph{2019 IEEE Intelligent vehicles
  symposium (IV)}}. IEEE, \bibinfo{pages}{438--445}.
\newblock


\bibitem[\protect\citeauthoryear{Borg}{Borg}{2021}]%
        {borg_AIQ_2021}
\bibfield{author}{\bibinfo{person}{Markus Borg}.}
  \bibinfo{year}{2021}\natexlab{}.
\newblock \showarticletitle{The AIQ Meta-Testbed: Pragmatically Bridging
  Academic AI Testing and Industrial Q Needs}. In
  \bibinfo{booktitle}{\emph{Software Quality: Future Perspectives on Software
  Engineering Quality}}, \bibfield{editor}{\bibinfo{person}{Dietmar Winkler},
  \bibinfo{person}{Stefan Biffl}, \bibinfo{person}{Daniel Mendez},
  \bibinfo{person}{Manuel Wimmer}, {and} \bibinfo{person}{Johannes Bergsmann}}
  (Eds.). \bibinfo{publisher}{Springer International Publishing},
  \bibinfo{address}{Cham}, \bibinfo{pages}{66--77}.
\newblock
\showISBNx{978-3-030-65854-0}


\bibitem[\protect\citeauthoryear{Borg}{Borg}{2022}]%
        {borg2022agility}
\bibfield{author}{\bibinfo{person}{Markus Borg}.}
  \bibinfo{year}{2022}\natexlab{}.
\newblock \showarticletitle{Agility in Software 2.0 -- Notebook Interfaces and
  MLOps with Buttresses and Rebars}. In \bibinfo{booktitle}{\emph{Proc. of the
  International Conference on Lean and Agile Software Development}}. Springer,
  \bibinfo{pages}{3--16}.
\newblock


\bibitem[\protect\citeauthoryear{Borg, Bronson, Christensson, Olsson,
  Lennartsson, Sonnsj{\"o}, Ebadi, and Karsberg}{Borg et~al\mbox{.}}{2021a}]%
        {borg2021exploring}
\bibfield{author}{\bibinfo{person}{Markus Borg}, \bibinfo{person}{Joshua
  Bronson}, \bibinfo{person}{Linus Christensson}, \bibinfo{person}{Fredrik
  Olsson}, \bibinfo{person}{Olof Lennartsson}, \bibinfo{person}{Elias
  Sonnsj{\"o}}, \bibinfo{person}{Hamid Ebadi}, {and} \bibinfo{person}{Martin
  Karsberg}.} \bibinfo{year}{2021}\natexlab{a}.
\newblock \showarticletitle{Exploring the Assessment List for Trustworthy AI in
  the Context of Advanced Driver-Assistance Systems}. In
  \bibinfo{booktitle}{\emph{2021 IEEE/ACM 2nd International Workshop on Ethics
  in Software Engineering Research and Practice (SEthics)}}. IEEE,
  \bibinfo{pages}{5--12}.
\newblock


\bibitem[\protect\citeauthoryear{Borg, Englund, Wnuk, Duran, Lewandowski, Gao,
  Tan, Kaijser, L{\"o}nn, and T{\"o}rnqvist}{Borg et~al\mbox{.}}{2019}]%
        {borg2019safely}
\bibfield{author}{\bibinfo{person}{Markus Borg}, \bibinfo{person}{Cristofer
  Englund}, \bibinfo{person}{Krzysztof Wnuk}, \bibinfo{person}{Boris Duran},
  \bibinfo{person}{Christoffer Lewandowski}, \bibinfo{person}{Shenjian Gao},
  \bibinfo{person}{Yanwen Tan}, \bibinfo{person}{Henrik Kaijser},
  \bibinfo{person}{Henrik L{\"o}nn}, {and} \bibinfo{person}{Jonnas
  T{\"o}rnqvist}.} \bibinfo{year}{2019}\natexlab{}.
\newblock \showarticletitle{Safely Entering the Deep: A Review of Verification
  and Validation for Machine Learning and a Challenge Elicitation in the
  Automotive Industry}.
\newblock \bibinfo{journal}{\emph{Journal of Automotive Software Engineering}}
  \bibinfo{volume}{1}, \bibinfo{number}{1} (\bibinfo{year}{2019}),
  \bibinfo{pages}{1--19}.
\newblock


\bibitem[\protect\citeauthoryear{Borg, Jabangwe, {\AA}berg, Ekblom, Hedlund,
  and Lidfeldt}{Borg et~al\mbox{.}}{2021b}]%
        {borg2021test}
\bibfield{author}{\bibinfo{person}{Markus Borg}, \bibinfo{person}{Ronald
  Jabangwe}, \bibinfo{person}{Simon {\AA}berg}, \bibinfo{person}{Arvid Ekblom},
  \bibinfo{person}{Ludwig Hedlund}, {and} \bibinfo{person}{August Lidfeldt}.}
  \bibinfo{year}{2021}\natexlab{b}.
\newblock \showarticletitle{Test automation with grad-CAM Heatmaps-A future
  pipe segment in MLOps for Vision AI?}. In \bibinfo{booktitle}{\emph{2021 IEEE
  International Conference on Software Testing, Verification and Validation
  Workshops (ICSTW)}}. IEEE, \bibinfo{pages}{175--181}.
\newblock


\bibitem[\protect\citeauthoryear{Bosch, Olsson, and Crnkovic}{Bosch
  et~al\mbox{.}}{2021}]%
        {bosch2021engineering}
\bibfield{author}{\bibinfo{person}{Jan Bosch},
  \bibinfo{person}{Helena~Holmstr{\"o}m Olsson}, {and} \bibinfo{person}{Ivica
  Crnkovic}.} \bibinfo{year}{2021}\natexlab{}.
\newblock \showarticletitle{Engineering ai systems: A research agenda}.
\newblock In \bibinfo{booktitle}{\emph{Artificial Intelligence Paradigms for
  Smart Cyber-Physical Systems}}. \bibinfo{publisher}{IGI Global},
  \bibinfo{pages}{1--19}.
\newblock


\bibitem[\protect\citeauthoryear{Breck, Polyzotis, Roy, Whang, and
  Zinkevich}{Breck et~al\mbox{.}}{2019}]%
        {breck2019data}
\bibfield{author}{\bibinfo{person}{Eric Breck}, \bibinfo{person}{Neoklis
  Polyzotis}, \bibinfo{person}{Sudip Roy}, \bibinfo{person}{Steven Whang},
  {and} \bibinfo{person}{Martin Zinkevich}.} \bibinfo{year}{2019}\natexlab{}.
\newblock \showarticletitle{Data Validation for Machine Learning.}. In
  \bibinfo{booktitle}{\emph{MLSys}}.
\newblock


\bibitem[\protect\citeauthoryear{Byun, Sharma, Vijayakumar, Rayadurgam, and
  Cofer}{Byun et~al\mbox{.}}{2019}]%
        {byun2019input}
\bibfield{author}{\bibinfo{person}{Taejoon Byun}, \bibinfo{person}{Vaibhav
  Sharma}, \bibinfo{person}{Abhishek Vijayakumar}, \bibinfo{person}{Sanjai
  Rayadurgam}, {and} \bibinfo{person}{Darren Cofer}.}
  \bibinfo{year}{2019}\natexlab{}.
\newblock \showarticletitle{Input prioritization for testing neural networks}.
  In \bibinfo{booktitle}{\emph{2019 IEEE International Conference On Artificial
  Intelligence Testing (AITest)}}. IEEE, \bibinfo{pages}{63--70}.
\newblock


\bibitem[\protect\citeauthoryear{Carleton, Harper, Lyu, Eldh, Xie, and
  Menzies}{Carleton et~al\mbox{.}}{2020}]%
        {carleton2020expert}
\bibfield{author}{\bibinfo{person}{Anita~D Carleton}, \bibinfo{person}{Erin
  Harper}, \bibinfo{person}{Michael~R Lyu}, \bibinfo{person}{Sigrid Eldh},
  \bibinfo{person}{Tao Xie}, {and} \bibinfo{person}{Tim Menzies}.}
  \bibinfo{year}{2020}\natexlab{}.
\newblock \showarticletitle{Expert Perspectives on AI}.
\newblock \bibinfo{journal}{\emph{IEEE Software}} \bibinfo{volume}{37},
  \bibinfo{number}{4} (\bibinfo{year}{2020}), \bibinfo{pages}{87--94}.
\newblock


\bibitem[\protect\citeauthoryear{Cartaxo, Pinto, and Soares}{Cartaxo
  et~al\mbox{.}}{2020}]%
        {Cartaxo2020}
\bibfield{author}{\bibinfo{person}{Bruno Cartaxo}, \bibinfo{person}{Gustavo
  Pinto}, {and} \bibinfo{person}{Sergio Soares}.}
  \bibinfo{year}{2020}\natexlab{}.
\newblock \bibinfo{booktitle}{\emph{Rapid Reviews in Software Engineering}}.
\newblock \bibinfo{publisher}{Springer International Publishing},
  \bibinfo{address}{Cham}, \bibinfo{pages}{357--384}.
\newblock
\showISBNx{978-3-030-32489-6}
\urldef\tempurl%
\url{https://doi.org/10.1007/978-3-030-32489-6_13}
\showDOI{\tempurl}


\bibitem[\protect\citeauthoryear{Cheng, Huang, Ruess, Yasuoka,
  et~al\mbox{.}}{Cheng et~al\mbox{.}}{2018}]%
        {cheng2018towards}
\bibfield{author}{\bibinfo{person}{Chih-Hong Cheng}, \bibinfo{person}{Chung-Hao
  Huang}, \bibinfo{person}{Harald Ruess}, \bibinfo{person}{Hirotoshi Yasuoka},
  {et~al\mbox{.}}} \bibinfo{year}{2018}\natexlab{}.
\newblock \showarticletitle{Towards dependability metrics for neural networks}.
  In \bibinfo{booktitle}{\emph{2018 16th ACM/IEEE International Conference on
  Formal Methods and Models for System Design (MEMOCODE)}}. IEEE,
  \bibinfo{pages}{1--4}.
\newblock


\bibitem[\protect\citeauthoryear{Ebadi, Moghadam, Borg, Gay, Fontes, and
  Socha}{Ebadi et~al\mbox{.}}{2021}]%
        {ebadi2021efficient}
\bibfield{author}{\bibinfo{person}{Hamid Ebadi},
  \bibinfo{person}{Mahshid~Helali Moghadam}, \bibinfo{person}{Markus Borg},
  \bibinfo{person}{Gregory Gay}, \bibinfo{person}{Afonso Fontes}, {and}
  \bibinfo{person}{Kasper Socha}.} \bibinfo{year}{2021}\natexlab{}.
\newblock \showarticletitle{Efficient and Effective Generation of Test Cases
  for Pedestrian Detection-Search-based Software Testing of Baidu Apollo in
  SVL}. In \bibinfo{booktitle}{\emph{2021 IEEE International Conference on
  Artificial Intelligence Testing (AITest)}}. IEEE, \bibinfo{pages}{103--110}.
\newblock


\bibitem[\protect\citeauthoryear{Engstr\"{o}m, Petersen, Ali, and
  Bjarnason}{Engstr\"{o}m et~al\mbox{.}}{2017}]%
        {engstrom_SERP-test_2017}
\bibfield{author}{\bibinfo{person}{Emelie Engstr\"{o}m}, \bibinfo{person}{Kai
  Petersen}, \bibinfo{person}{Nauman~Bin Ali}, {and} \bibinfo{person}{Elizabeth
  Bjarnason}.} \bibinfo{year}{2017}\natexlab{}.
\newblock \showarticletitle{SERP-Test: A Taxonomy for Supporting
  Industry---Academia Communication}.
\newblock \bibinfo{journal}{\emph{Software Quality Journal}}
  \bibinfo{volume}{25}, \bibinfo{number}{4} (\bibinfo{date}{dec}
  \bibinfo{year}{2017}), \bibinfo{pages}{1269–1305}.
\newblock
\showISSN{0963-9314}
\urldef\tempurl%
\url{https://doi.org/10.1007/s11219-016-9322-x}
\showDOI{\tempurl}


\bibitem[\protect\citeauthoryear{Felderer, Russo, and Auer}{Felderer
  et~al\mbox{.}}{2019}]%
        {felderer2019testing}
\bibfield{author}{\bibinfo{person}{Michael Felderer}, \bibinfo{person}{Barbara
  Russo}, {and} \bibinfo{person}{Florian Auer}.}
  \bibinfo{year}{2019}\natexlab{}.
\newblock \showarticletitle{On testing data-intensive software systems}.
\newblock In \bibinfo{booktitle}{\emph{Security and Quality in Cyber-Physical
  Systems Engineering}}. \bibinfo{publisher}{Springer},
  \bibinfo{pages}{129--148}.
\newblock


\bibitem[\protect\citeauthoryear{Henriksson, Berger, Borg, Tornberg, Englund,
  Sathyamoorthy, and Ursing}{Henriksson et~al\mbox{.}}{2019}]%
        {henriksson2019towards}
\bibfield{author}{\bibinfo{person}{Jens Henriksson}, \bibinfo{person}{Christian
  Berger}, \bibinfo{person}{Markus Borg}, \bibinfo{person}{Lars Tornberg},
  \bibinfo{person}{Cristofer Englund}, \bibinfo{person}{Sankar~Raman
  Sathyamoorthy}, {and} \bibinfo{person}{Stig Ursing}.}
  \bibinfo{year}{2019}\natexlab{}.
\newblock \showarticletitle{Towards structured evaluation of deep neural
  network supervisors}. In \bibinfo{booktitle}{\emph{2019 IEEE International
  Conference On Artificial Intelligence Testing (AITest)}}. IEEE,
  \bibinfo{pages}{27--34}.
\newblock


\bibitem[\protect\citeauthoryear{Hynes, Sculley, and Terry}{Hynes
  et~al\mbox{.}}{2017}]%
        {hynes2017data}
\bibfield{author}{\bibinfo{person}{Nick Hynes}, \bibinfo{person}{D Sculley},
  {and} \bibinfo{person}{Michael Terry}.} \bibinfo{year}{2017}\natexlab{}.
\newblock \showarticletitle{The data linter: Lightweight, automated sanity
  checking for ml data sets}. In \bibinfo{booktitle}{\emph{NIPS MLSys
  Workshop}}.
\newblock


\bibitem[\protect\citeauthoryear{ISO/IEC}{ISO/IEC}{2008}]%
        {iso25012}
\bibfield{author}{\bibinfo{person}{ISO/IEC}.} \bibinfo{year}{2008}\natexlab{}.
\newblock \bibinfo{title}{ISO 25012 Systems and software engineering –
  Systems and software quality requirements and evaluation (SQuaRE) - Data
  quality model}.
\newblock
\newblock


\bibitem[\protect\citeauthoryear{ISO/IEC}{ISO/IEC}{2011}]%
        {iso25010}
\bibfield{author}{\bibinfo{person}{ISO/IEC}.} \bibinfo{year}{2011}\natexlab{}.
\newblock \bibinfo{title}{ISO 25010 Systems and software engineering –
  Systems and software quality requirements and evaluation (SQuaRE) - System
  and software quality models}.
\newblock
\newblock


\bibitem[\protect\citeauthoryear{Jiang, Jiang, Zhi, Dong, Li, Ma, Wang, Dong,
  Shen, and Wang}{Jiang et~al\mbox{.}}{2017}]%
        {jiang2017artificial}
\bibfield{author}{\bibinfo{person}{Fei Jiang}, \bibinfo{person}{Yong Jiang},
  \bibinfo{person}{Hui Zhi}, \bibinfo{person}{Yi Dong}, \bibinfo{person}{Hao
  Li}, \bibinfo{person}{Sufeng Ma}, \bibinfo{person}{Yilong Wang},
  \bibinfo{person}{Qiang Dong}, \bibinfo{person}{Haipeng Shen}, {and}
  \bibinfo{person}{Yongjun Wang}.} \bibinfo{year}{2017}\natexlab{}.
\newblock \showarticletitle{Artificial intelligence in healthcare: past,
  present and future}.
\newblock \bibinfo{journal}{\emph{Stroke and vascular neurology}}
  \bibinfo{volume}{2}, \bibinfo{number}{4} (\bibinfo{year}{2017}).
\newblock


\bibitem[\protect\citeauthoryear{Kim, Feldt, and Yoo}{Kim
  et~al\mbox{.}}{2019}]%
        {kim2019guiding}
\bibfield{author}{\bibinfo{person}{Jinhan Kim}, \bibinfo{person}{Robert Feldt},
  {and} \bibinfo{person}{Shin Yoo}.} \bibinfo{year}{2019}\natexlab{}.
\newblock \showarticletitle{Guiding deep learning system testing using surprise
  adequacy}. In \bibinfo{booktitle}{\emph{2019 IEEE/ACM 41st International
  Conference on Software Engineering (ICSE)}}. IEEE,
  \bibinfo{pages}{1039--1049}.
\newblock


\bibitem[\protect\citeauthoryear{Kim, Ju, Feldt, and Yoo}{Kim
  et~al\mbox{.}}{2020}]%
        {kim2020reducing}
\bibfield{author}{\bibinfo{person}{Jinhan Kim}, \bibinfo{person}{Jeongil Ju},
  \bibinfo{person}{Robert Feldt}, {and} \bibinfo{person}{Shin Yoo}.}
  \bibinfo{year}{2020}\natexlab{}.
\newblock \showarticletitle{Reducing dnn labelling cost using surprise
  adequacy: An industrial case study for autonomous driving}. In
  \bibinfo{booktitle}{\emph{Proceedings of the 28th ACM Joint Meeting on
  European Software Engineering Conference and Symposium on the Foundations of
  Software Engineering}}. \bibinfo{pages}{1466--1476}.
\newblock


\bibitem[\protect\citeauthoryear{Krishnan, Wang, Wu, Franklin, and
  Goldberg}{Krishnan et~al\mbox{.}}{2016}]%
        {krishnan2016activeclean}
\bibfield{author}{\bibinfo{person}{Sanjay Krishnan}, \bibinfo{person}{Jiannan
  Wang}, \bibinfo{person}{Eugene Wu}, \bibinfo{person}{Michael~J. Franklin},
  {and} \bibinfo{person}{Ken Goldberg}.} \bibinfo{year}{2016}\natexlab{}.
\newblock \showarticletitle{ActiveClean: Interactive Data Cleaning for
  Statistical Modeling}.
\newblock \bibinfo{journal}{\emph{Proc. VLDB Endow.}} \bibinfo{volume}{9},
  \bibinfo{number}{12} (\bibinfo{date}{aug} \bibinfo{year}{2016}),
  \bibinfo{pages}{948–959}.
\newblock
\showISSN{2150-8097}


\bibitem[\protect\citeauthoryear{Lwakatare, Raj, Crnkovic, Bosch, and
  Olsson}{Lwakatare et~al\mbox{.}}{2020}]%
        {LWAKATARE2020106368}
\bibfield{author}{\bibinfo{person}{Lucy~Ellen Lwakatare},
  \bibinfo{person}{Aiswarya Raj}, \bibinfo{person}{Ivica Crnkovic},
  \bibinfo{person}{Jan Bosch}, {and} \bibinfo{person}{Helena~Holmström
  Olsson}.} \bibinfo{year}{2020}\natexlab{}.
\newblock \showarticletitle{Large-scale machine learning systems in real-world
  industrial settings: A review of challenges and solutions}.
\newblock \bibinfo{journal}{\emph{Information and Software Technology}}
  \bibinfo{volume}{127} (\bibinfo{year}{2020}), \bibinfo{pages}{106368}.
\newblock
\showISSN{0950-5849}
\urldef\tempurl%
\url{https://doi.org/10.1016/j.infsof.2020.106368}
\showDOI{\tempurl}


\bibitem[\protect\citeauthoryear{Ma, Zhang, Sun, Xue, Li, Juefei-Xu, Xie, Li,
  Liu, Zhao, et~al\mbox{.}}{Ma et~al\mbox{.}}{2018}]%
        {ma2018deepmutation}
\bibfield{author}{\bibinfo{person}{Lei Ma}, \bibinfo{person}{Fuyuan Zhang},
  \bibinfo{person}{Jiyuan Sun}, \bibinfo{person}{Minhui Xue},
  \bibinfo{person}{Bo Li}, \bibinfo{person}{Felix Juefei-Xu},
  \bibinfo{person}{Chao Xie}, \bibinfo{person}{Li Li}, \bibinfo{person}{Yang
  Liu}, \bibinfo{person}{Jianjun Zhao}, {et~al\mbox{.}}}
  \bibinfo{year}{2018}\natexlab{}.
\newblock \showarticletitle{Deepmutation: Mutation testing of deep learning
  systems}. In \bibinfo{booktitle}{\emph{2018 IEEE 29th International Symposium
  on Software Reliability Engineering (ISSRE)}}. IEEE,
  \bibinfo{pages}{100--111}.
\newblock


\bibitem[\protect\citeauthoryear{Marijan, Gotlieb, and Kumar~Ahuja}{Marijan
  et~al\mbox{.}}{2019}]%
        {8718214}
\bibfield{author}{\bibinfo{person}{Dusica Marijan}, \bibinfo{person}{Arnaud
  Gotlieb}, {and} \bibinfo{person}{Mohit Kumar~Ahuja}.}
  \bibinfo{year}{2019}\natexlab{}.
\newblock \showarticletitle{Challenges of Testing Machine Learning Based
  Systems}. In \bibinfo{booktitle}{\emph{2019 IEEE International Conference On
  Artificial Intelligence Testing (AITest)}}. \bibinfo{pages}{101--102}.
\newblock
\urldef\tempurl%
\url{https://doi.org/10.1109/AITest.2019.00010}
\showDOI{\tempurl}


\bibitem[\protect\citeauthoryear{Moghadam, Borg, and Mousavirad}{Moghadam
  et~al\mbox{.}}{2021}]%
        {moghadam2021deeper}
\bibfield{author}{\bibinfo{person}{Mahshid~Helali Moghadam},
  \bibinfo{person}{Markus Borg}, {and} \bibinfo{person}{Seyed~Jalaleddin
  Mousavirad}.} \bibinfo{year}{2021}\natexlab{}.
\newblock \showarticletitle{Deeper at the sbst 2021 tool competition: ADAS
  testing using multi-objective search}. In \bibinfo{booktitle}{\emph{2021
  IEEE/ACM 14th International Workshop on Search-Based Software Testing
  (SBST)}}. IEEE, \bibinfo{pages}{40--41}.
\newblock


\bibitem[\protect\citeauthoryear{Petersen and Engstr\"{o}m}{Petersen and
  Engstr\"{o}m}{2014}]%
        {petersenSERP_2014}
\bibfield{author}{\bibinfo{person}{Kai Petersen} {and} \bibinfo{person}{Emelie
  Engstr\"{o}m}.} \bibinfo{year}{2014}\natexlab{}.
\newblock \showarticletitle{Finding Relevant Research Solutions for Practical
  Problems: The Serp Taxonomy Architecture}. In
  \bibinfo{booktitle}{\emph{Proceedings of the 2014 International Workshop on
  Long-Term Industrial Collaboration on Software Engineering}}
  \emph{(\bibinfo{series}{WISE '14})}. \bibinfo{publisher}{Association for
  Computing Machinery}, \bibinfo{address}{New York, NY, USA},
  \bibinfo{pages}{13–20}.
\newblock
\showISBNx{9781450330459}
\urldef\tempurl%
\url{https://doi.org/10.1145/2647648.2647650}
\showDOI{\tempurl}


\bibitem[\protect\citeauthoryear{Pleiss, Zhang, Elenberg, and
  Weinberger}{Pleiss et~al\mbox{.}}{2020}]%
        {NEURIPS2020_c6102b37}
\bibfield{author}{\bibinfo{person}{Geoff Pleiss}, \bibinfo{person}{Tianyi
  Zhang}, \bibinfo{person}{Ethan Elenberg}, {and} \bibinfo{person}{Kilian~Q
  Weinberger}.} \bibinfo{year}{2020}\natexlab{}.
\newblock \showarticletitle{Identifying Mislabeled Data using the Area Under
  the Margin Ranking}. In \bibinfo{booktitle}{\emph{Advances in Neural
  Information Processing Systems}},
  \bibfield{editor}{\bibinfo{person}{H.~Larochelle},
  \bibinfo{person}{M.~Ranzato}, \bibinfo{person}{R.~Hadsell},
  \bibinfo{person}{M.~F. Balcan}, {and} \bibinfo{person}{H.~Lin}} (Eds.),
  Vol.~\bibinfo{volume}{33}. \bibinfo{pages}{17044--17056}.
\newblock


\bibitem[\protect\citeauthoryear{Riccio, Jahangirova, Stocco, Humbatova, Weiss,
  and Tonella}{Riccio et~al\mbox{.}}{2020}]%
        {riccio2020testing}
\bibfield{author}{\bibinfo{person}{Vincenzo Riccio}, \bibinfo{person}{Gunel
  Jahangirova}, \bibinfo{person}{Andrea Stocco}, \bibinfo{person}{Nargiz
  Humbatova}, \bibinfo{person}{Michael Weiss}, {and} \bibinfo{person}{Paolo
  Tonella}.} \bibinfo{year}{2020}\natexlab{}.
\newblock \showarticletitle{Testing machine learning based systems: a
  systematic mapping}.
\newblock \bibinfo{journal}{\emph{Empirical Software Engineering}}
  \bibinfo{volume}{25}, \bibinfo{number}{6} (\bibinfo{year}{2020}),
  \bibinfo{pages}{5193--5254}.
\newblock


\bibitem[\protect\citeauthoryear{Rico, Ali, Engstr{\"o}m, and H{\"o}st}{Rico
  et~al\mbox{.}}{2020}]%
        {Rico2020GuidelinesFC}
\bibfield{author}{\bibinfo{person}{Sergio Rico}, \bibinfo{person}{N. Ali},
  \bibinfo{person}{Emelie Engstr{\"o}m}, {and} \bibinfo{person}{Martin
  H{\"o}st}.} \bibinfo{year}{2020}\natexlab{}.
\newblock \showarticletitle{Guidelines for conducting interactive rapid reviews
  in software engineering -- from a focus on technology transfer to knowledge
  exchange}.
\newblock


\bibitem[\protect\citeauthoryear{Schelter, Lange, Schmidt, Celikel, Biessmann,
  and Grafberger}{Schelter et~al\mbox{.}}{2018}]%
        {schelter2018automating}
\bibfield{author}{\bibinfo{person}{Sebastian Schelter}, \bibinfo{person}{Dustin
  Lange}, \bibinfo{person}{Philipp Schmidt}, \bibinfo{person}{Meltem Celikel},
  \bibinfo{person}{Felix Biessmann}, {and} \bibinfo{person}{Andreas
  Grafberger}.} \bibinfo{year}{2018}\natexlab{}.
\newblock \showarticletitle{Automating large-scale data quality verification}.
\newblock \bibinfo{journal}{\emph{Proceedings of the VLDB Endowment}}
  \bibinfo{volume}{11}, \bibinfo{number}{12} (\bibinfo{year}{2018}),
  \bibinfo{pages}{1781--1794}.
\newblock


\bibitem[\protect\citeauthoryear{Sculley et~al\mbox{.}}{Sculley
  et~al\mbox{.}}{2015}]%
        {sculley_hidden_2015}
\bibfield{author}{\bibinfo{person}{D. Sculley} {et~al\mbox{.}}}
  \bibinfo{year}{2015}\natexlab{}.
\newblock \showarticletitle{Hidden {Technical} {Debt} in {Machine} {Learning}
  {Systems}}. In \bibinfo{booktitle}{\emph{Proc. of the 28th {Int'l} {Conf.} on
  {Neural} {Information} {Proc.} {Systems}}}. \bibinfo{pages}{2503--2511}.
\newblock


\bibitem[\protect\citeauthoryear{Sherin, khan, and Iqbal}{Sherin
  et~al\mbox{.}}{2019}]%
        {sherin2019systematic}
\bibfield{author}{\bibinfo{person}{Salman Sherin},
  \bibinfo{person}{Muhammad~Uzair khan}, {and} \bibinfo{person}{Muhammad~Zohaib
  Iqbal}.} \bibinfo{year}{2019}\natexlab{}.
\newblock \bibinfo{title}{A Systematic Mapping Study on Testing of Machine
  Learning Programs}.
\newblock
\newblock
\showeprint[arxiv]{cs.LG/1907.09427}


\bibitem[\protect\citeauthoryear{Song, Borg, Engström, Ardö, and Rico}{Song
  et~al\mbox{.}}{2022}]%
        {qunying_song_2022_5865070}
\bibfield{author}{\bibinfo{person}{Qunying Song}, \bibinfo{person}{Markus
  Borg}, \bibinfo{person}{Emelie Engström}, \bibinfo{person}{Håkan Ardö},
  {and} \bibinfo{person}{Sergio Rico}.} \bibinfo{year}{2022}\natexlab{}.
\newblock \bibinfo{title}{Primary Studies}.
\newblock
\newblock
\urldef\tempurl%
\url{https://doi.org/10.5281/zenodo.5865070}
\showDOI{\tempurl}


\bibitem[\protect\citeauthoryear{Storey, Engstrom, Höst, Runeson, and
  Bjarnason}{Storey et~al\mbox{.}}{2017}]%
        {storey_2017_using}
\bibfield{author}{\bibinfo{person}{Margaret-Anne Storey},
  \bibinfo{person}{Emelie Engstrom}, \bibinfo{person}{Martin Höst},
  \bibinfo{person}{Per Runeson}, {and} \bibinfo{person}{Elizabeth Bjarnason}.}
  \bibinfo{year}{2017}\natexlab{}.
\newblock \showarticletitle{Using a Visual Abstract as a Lens for Communicating
  and Promoting Design Science Research in Software Engineering}. In
  \bibinfo{booktitle}{\emph{2017 ACM/IEEE International Symposium on Empirical
  Software Engineering and Measurement (ESEM)}}. \bibinfo{pages}{181--186}.
\newblock
\urldef\tempurl%
\url{https://doi.org/10.1109/ESEM.2017.28}
\showDOI{\tempurl}


\bibitem[\protect\citeauthoryear{Tian, Pei, Jana, and Ray}{Tian
  et~al\mbox{.}}{2018}]%
        {tian2018deeptest}
\bibfield{author}{\bibinfo{person}{Yuchi Tian}, \bibinfo{person}{Kexin Pei},
  \bibinfo{person}{Suman Jana}, {and} \bibinfo{person}{Baishakhi Ray}.}
  \bibinfo{year}{2018}\natexlab{}.
\newblock \showarticletitle{Deeptest: Automated testing of
  deep-neural-network-driven autonomous cars}. In
  \bibinfo{booktitle}{\emph{Proceedings of the 40th international conference on
  software engineering}}. \bibinfo{pages}{303--314}.
\newblock


\bibitem[\protect\citeauthoryear{Tricco, Garritty, Boulos, Lockwood, Wilson,
  McGowan, McCaul, Hutton, Clement, Mittmann, et~al\mbox{.}}{Tricco
  et~al\mbox{.}}{2020}]%
        {tricco2020rapid}
\bibfield{author}{\bibinfo{person}{Andrea~C Tricco},
  \bibinfo{person}{Chantelle~M Garritty}, \bibinfo{person}{Leah Boulos},
  \bibinfo{person}{Craig Lockwood}, \bibinfo{person}{Michael Wilson},
  \bibinfo{person}{Jessie McGowan}, \bibinfo{person}{Michael McCaul},
  \bibinfo{person}{Brian Hutton}, \bibinfo{person}{Fiona Clement},
  \bibinfo{person}{Nicole Mittmann}, {et~al\mbox{.}}}
  \bibinfo{year}{2020}\natexlab{}.
\newblock \showarticletitle{Rapid review methods more challenging during
  COVID-19: commentary with a focus on 8 knowledge synthesis steps}.
\newblock \bibinfo{journal}{\emph{Journal of clinical epidemiology}}
  \bibinfo{volume}{126} (\bibinfo{year}{2020}), \bibinfo{pages}{177--183}.
\newblock


\bibitem[\protect\citeauthoryear{Uesato, Kumar, Szepesvari, Erez, Ruderman,
  Anderson, Heess, Kohli, et~al\mbox{.}}{Uesato et~al\mbox{.}}{2018}]%
        {uesato2018rigorous}
\bibfield{author}{\bibinfo{person}{Jonathan Uesato}, \bibinfo{person}{Ananya
  Kumar}, \bibinfo{person}{Csaba Szepesvari}, \bibinfo{person}{Tom Erez},
  \bibinfo{person}{Avraham Ruderman}, \bibinfo{person}{Keith Anderson},
  \bibinfo{person}{Nicolas Heess}, \bibinfo{person}{Pushmeet Kohli},
  {et~al\mbox{.}}} \bibinfo{year}{2018}\natexlab{}.
\newblock \showarticletitle{Rigorous agent evaluation: An adversarial approach
  to uncover catastrophic failures}.
\newblock \bibinfo{journal}{\emph{arXiv preprint arXiv:1812.01647}}
  (\bibinfo{year}{2018}).
\newblock


\bibitem[\protect\citeauthoryear{Vidot, Gabreau, Ober, and Ober}{Vidot
  et~al\mbox{.}}{2021}]%
        {vidot2021certification}
\bibfield{author}{\bibinfo{person}{Guillaume Vidot},
  \bibinfo{person}{Christophe Gabreau}, \bibinfo{person}{Ileana Ober}, {and}
  \bibinfo{person}{Iulian Ober}.} \bibinfo{year}{2021}\natexlab{}.
\newblock \showarticletitle{Certification of embedded systems based on Machine
  Learning: A survey}.
\newblock \bibinfo{journal}{\emph{arXiv preprint arXiv:2106.07221}}
  (\bibinfo{year}{2021}).
\newblock


\bibitem[\protect\citeauthoryear{Vogelsang and Borg}{Vogelsang and
  Borg}{2019}]%
        {vogelsang_requirements_2019}
\bibfield{author}{\bibinfo{person}{Andreas Vogelsang} {and}
  \bibinfo{person}{Markus Borg}.} \bibinfo{year}{2019}\natexlab{}.
\newblock \showarticletitle{Requirements {Engineering} for {Machine}
  {Learning}: {Perspectives} from {Data} {Scientists}}. In
  \bibinfo{booktitle}{\emph{Proc. of the 27th {Int'l} {Requirements}
  {Engineering} {Conf.} {Workshops}}}. \bibinfo{pages}{245--251}.
\newblock
\urldef\tempurl%
\url{https://doi.org/10.1109/REW.2019.00050}
\showDOI{\tempurl}


\bibitem[\protect\citeauthoryear{Zhang, Barr, Guedj, Harman, and
  Shawe-Taylor}{Zhang et~al\mbox{.}}{2019}]%
        {zhang2019perturbed}
\bibfield{author}{\bibinfo{person}{Jie Zhang}, \bibinfo{person}{Earl Barr},
  \bibinfo{person}{Benjamin Guedj}, \bibinfo{person}{Mark Harman}, {and}
  \bibinfo{person}{John Shawe-Taylor}.} \bibinfo{year}{2019}\natexlab{}.
\newblock \showarticletitle{Perturbed model validation: A new framework to
  validate model relevance}.
\newblock  (\bibinfo{year}{2019}).
\newblock


\bibitem[\protect\citeauthoryear{Zhang, Harman, Ma, and Liu}{Zhang
  et~al\mbox{.}}{2020}]%
        {zhang2020machine}
\bibfield{author}{\bibinfo{person}{Jie~M Zhang}, \bibinfo{person}{Mark Harman},
  \bibinfo{person}{Lei Ma}, {and} \bibinfo{person}{Yang Liu}.}
  \bibinfo{year}{2020}\natexlab{}.
\newblock \showarticletitle{Machine learning testing: Survey, landscapes and
  horizons}.
\newblock \bibinfo{journal}{\emph{IEEE Transactions on Software Engineering}}
  (\bibinfo{year}{2020}).
\newblock


\bibitem[\protect\citeauthoryear{Zhang, Zhang, Zhang, Liu, and Khurshid}{Zhang
  et~al\mbox{.}}{2018}]%
        {zhang2018deeproad}
\bibfield{author}{\bibinfo{person}{Mengshi Zhang}, \bibinfo{person}{Yuqun
  Zhang}, \bibinfo{person}{Lingming Zhang}, \bibinfo{person}{Cong Liu}, {and}
  \bibinfo{person}{Sarfraz Khurshid}.} \bibinfo{year}{2018}\natexlab{}.
\newblock \showarticletitle{DeepRoad: GAN-based metamorphic testing and input
  validation framework for autonomous driving systems}. In
  \bibinfo{booktitle}{\emph{2018 33rd IEEE/ACM International Conference on
  Automated Software Engineering (ASE)}}. IEEE, \bibinfo{pages}{132--142}.
\newblock


\end{thebibliography}
